\documentclass[journal,10pt]{IEEEtran}
\usepackage{cite,graphicx,amssymb,amsmath,color,textcomp,tabularx}
\usepackage{multicol}
\usepackage{graphicx}
\usepackage{epstopdf}
\usepackage{stfloats}
\usepackage{algorithm,algorithmic}

\newtheorem{theorem}{Theorem}

\newtheorem{remark}{Remark}
\newtheorem{corollary}{Corollary}
\newtheorem{proposition}{Proposition}
\IEEEoverridecommandlockouts

\begin{document}
\title{\LARGE{Energy-Efficient Transmission Design in Non-Orthogonal Multiple Access}}
\author
{Yi Zhang, \IEEEmembership{Student Member,~IEEE}, Hui-Ming Wang, \IEEEmembership{Senior Member,~IEEE}, Tong-Xing Zheng and Qian Yang\vspace{-16pt}
\thanks{Manuscript received March 27, 2016; accepted June 4, 2016. The associate editor coordinating the review of this paper and approving it for publication was Prof. Guoqiang Mao. The work was supported by the Fund for the Author of National Excellent Doctoral Dissertation of China under Grant 201340, the New Century Excellent Talents Support Fund of China under Grant NCET-13-0458 and by the Young Talent Support Fund of Science and Technology of Shaanxi Province under Grant 2015KJXX-01.

Copyright (c) 2015 IEEE. Personal use of this material is permitted. However, permission to use this material for any other purposes must be obtained from the IEEE by sending a request to pubs-permissions@ieee.org.

The authors are with the Ministry of Education Key Lab for Intelligent Networks and Network Security, Xi'an Jiaotong University, Xi'an 710049, Shaanxi, China (e-mail: yi.zhang.cn@outlook.com; xjbswhm@gmail.com; txzheng@stu.xjtu.edu.cn; qian-yang@outlook.com). The corresponding author is Hui-Ming Wang.}}
\markboth{IEEE Transactions on Vehicular Technology}{}
\maketitle

\begin{abstract}
Non-orthogonal multiple access (NOMA) is considered as a promising technology for improving the spectral efficiency (SE) in 5G. In this correspondence, we study the benefit of NOMA in enhancing energy efficiency (EE) for a multi-user downlink transmission, where the EE is defined as the ratio of the achievable sum rate of the users to the total power consumption. Our goal is to maximize the EE subject to a minimum required data rate for each user, which leads to a non-convex fractional programming problem. To solve it, we first establish the feasible range of the transmitting power that is able to support each user's data rate requirement. Then, we propose an EE-optimal power allocation strategy that maximizes the EE. Our numerical results show that NOMA has superior EE performance in comparison with conventional orthogonal multiple access (OMA).
\end{abstract}

\IEEEpeerreviewmaketitle

\vspace{-6pt}
\section{Introduction}
NOMA has been recognized as a promising candidate for 5G communication systems\cite{Noma_Performace}. In contrast with conventional OMA, e.g., time-division multiple access (TDMA), NOMA serves multiple users simultaneously via power domain division. Early literature on NOMA has mainly focused on the improvement of SE. For example, in \cite{NOMA_SISO_DING}, the authors analyzed the ergodic sum rate and the outage performance of a single-input single-output (SISO) NOMA system with randomly deployed users. In \cite{UserPairing}, the impact of user pairing on two-user SISO NOMA systems was considered. Besides, the power allocation among users in a SISO NOMA system was investigated in \cite{Fairness} from the perspective of user fairness.

In addition to SE, EE has recently drawn significant attention since the information and communication technology (ICT) accounts for around 5\% of the entire world energy consumption \cite{Pseudo-concavity}, which is becoming one of the major social and economical concerns worldwide. Currently, only a few works have studied NOMA from the perspective of EE. In \cite{EE2}, the EE optimization was performed in a fading multiple-input multiple-output (MIMO) NOMA system. However, the number of users is limited and fixed as two in \cite{EE2}, which greatly restrains the application of NOMA.

Motivated by the aforementioned observations, in this correspondence, we study the EE optimization in a downlink SISO NOMA system with multiple users, where each user has its own quality of service (QoS) requirement guaranteed by a minimum required data rate.
We first determine the minimum transmitting power that is able to support the required data rate for each user. Then an energy-efficient power allocation strategy is proposed to maximize the EE by solving a non-convex fractional programming problem. This optimization is further decoupled into two concatenate subproblems and solved one by one:
1) a non-convex multivariate optimization problem that is solved in closed form;
2) a strict pseudo-concave univariate optimization problem that is solved by the bisection method.
Our numerical results show that NOMA has superior EE performance compared with conventional OMA.

\section{System Model}
Consider a downlink transmission scenario wherein one single-antenna BS simultaneously serves $K$ single-antenna users. The channel from the BS to the $k$-th user, $1\leq k\leq K$, is modeled as $h_k=g_kd_k^{-\frac{\alpha}{2}}$, where $g_k$ is the Rayleigh fading coefficient, $d_k$ is the distance between the BS and the $k$-th user, and $\alpha$ is the path loss exponent. The instantaneous channel state information (CSI) of all users is known at the BS. Without loss of generality, we assume that the channel gains are sorted in the ascending order, i.e., $0<\left|h_1\right|^2\leq\left|h_2\right|^2...\leq\left|h_K\right|^2$.

According to the principle of NOMA\cite{Noma_Performace,NOMA_SISO_DING}, the BS broadcasts the superposition of $K$ signals to its $K$ users via power domain division. We denote $P$ as the total power available at the BS, $a_k$ as the $k$-th user's power allocation coefficient, which is defined as the ratio of the transmitting power for the $k$-th user's message to the total power $P$.
At receivers, successive interference cancellation (SIC) is used to eliminate the multi-user interference. Specifically, the $k$-th user first decodes the $i$-th user's message, $i<k$, and then removes this message from its received signal, in the order $i=1,2,...,k-1$; the messages for the $i$-th user, $i>k$, are treated as noise\cite{NOMA_SISO_DING}.
The achievable rate of the $k$-th user $R_k$ and the achievable sum rate of the system $R$ are given by
\begin{align}
    &R_k=\log_2\left(1+\frac{P\left|h_k\right|^2a_k}{P\left|h_k\right|^2\sum_{i=k+1}^Ka_i+\sigma^2}\right),\label{Rm}\\
    &R = \sum\nolimits_{k=1}^KR_k\label{R},
\end{align}
respectively\cite{NOMA_SISO_DING}, where $\sigma^2$ is the power of the additive noise.

\section{Problem Formulation}
As done in \cite{EE2,Pseudo-concavity}, the EE is defined as the ratio of the achievable sum rate of the system to the total power consumption, which is given by
$\textrm{EE}\triangleq\frac{R}{P_t+P_c}$,
where $P_t \triangleq \sum\nolimits_{k=1}^Ka_kP$ is the actually consumed transmitting power and $P_c$ is the constant power consumption of circuits.

Our design is based on providing QoS guarantees for all users. Each user has a minimum required data rate, denoted as $R_k^{\textrm{Min}}$ for $1\leq k\leq K$, i.e.,
\begin{equation}\label{RmQm}
    R_k \geq R_k^{\textrm{Min}}, ~~~1\leq k\leq K,
\end{equation}
which can be further transformed into
\begin{equation}\label{RmQm2}
a_k\geq A_k\left(\sum\nolimits_{i=k+1}^{K}a_i+\frac{\sigma^2}{P\left|h_{k}\right|^2}\right),~1\leq k\leq K,
\end{equation}
where $A_k\triangleq2^{R_k^{\textrm{Min}}}-1$. Thereby, the EE maximization problem is formulated as
\begin{subequations}\label{EE_MV_Original}
    \begin{align}
    &\max_{P_t,a_k,1\leq k\leq K}~\textrm{EE}\\
    &~~~~~~\textrm{s.t.}~~~~~~~P_t\leq P~~~\textrm{and}~~~P_t=\sum\nolimits_{k=1}^Ka_kP,\label{constraintP}\\
    &~~~~~~~~~~~~~~~~(\ref{RmQm2})\label{constraintQ}.
    \end{align}
\end{subequations}
Due to the minimum data rate constraints in (\ref{constraintQ}), problem (\ref{EE_MV_Original}) might be infeasible when the total power $P$ is not sufficiently large. Accordingly, there must exist a minimum transmitting power $P_{\textrm{Min}}$ that satisfies all users' data rate requirements and then problem (\ref{EE_MV_Original}) is feasible only under the condition $P\geq P_{\textrm{Min}}$. Thereby, it is important to firstly establish the feasible range of $P$, the derivation of which is discussed as follows.

\subsection{Minimum Required Transmitting Power $P_{\textrm{Min}}$}
Denote $P_k$ as the power allocated to the $k$-th user's message, then the problem of figuring out $P_{\textrm{Min}}$ is formulated as
\begin{subequations}\label{minp}
    \begin{align}
    &P_{\textrm{Min}}\triangleq\min_{P_k,1\leq k\leq K}~~\sum\nolimits_{k=1}^KP_k\label{Ob RmQm2} \\
    &\textrm{s.t.}~P_k\geq A_k\left(\sum\nolimits_{i=k+1}^{K}P_i+\frac{\sigma^2}{\left|h_{k}\right|^2}\right),~~~1\leq k\leq K,\label{C RmQm2}
    \end{align}
\end{subequations}
where (\ref{C RmQm2}) comes from the minimum data rate constraints in (\ref{RmQm2}). Problem (\ref{minp}) is solved by the following theorem.
\begin{theorem}\label{theo1}
The optimal solution to problem (\ref{minp}), denoted by $\{P_k^{\textrm{Min}}\}_{k=1}^{K}$, is given as
\begin{equation}\label{calculate Pkmin}
    P_k^{\textrm{Min}}=A_k\left(\sum\nolimits_{i=k+1}^{K}P_i^{\textrm{Min}}+\frac{\sigma^2}{\left|h_{k}\right|^2}\right),~~~1\leq k\leq K.
\end{equation}
\end{theorem}
\begin{IEEEproof}
It can be seen that problem (\ref{minp}) is convex, thus the following Karush-Kuhn-Tucker (KKT) conditions are necessary and sufficient for its optimal solution:
\begin{align}
    &1+\sum\nolimits_{i=1}^{k-1}\mu_iA_i =\mu_k,~~~~~~~~~~~~~~~~~~~~~~1\leq k\leq K,\label{KKT_Pmin}\\
    &\mu_k\left[A_k\left(\sum_{i=k+1}^{K}P_i+\frac{\sigma^2}{\left|h_{k}\right|^2}\right)-P_k\right]=0,~1\leq k\leq K,\\
    &\mu_k\geq 0,~~~~~~~~~~~~~~~~~~~~~~~~~~~~~~~~~~~~~~~~1\leq k\leq K,\label{mu_original_Pmin}
\end{align}
where $\{\mu_k\}_{k=1}^K$ are the Lagrange multipliers for the constraints in (\ref{C RmQm2}).
According to (\ref{KKT_Pmin}), we have $\mu_k> 0$ for $1\leq k\leq K$, because $\{A_k\}_{k=1}^{K}$ and $\{\mu_k\}_{k=1}^K$ are all nonnegative numbers. This indicates that the constraints in (\ref{C RmQm2}) are all satisfied at equality. Further, by setting the constraints in (\ref{C RmQm2}) to be active for $1\leq k\leq K$, the closed-form expressions of $\{P_k^{\textrm{Min}}\}_{k=1}^{K}$ are given by (\ref{calculate Pkmin}). Specifically, $\{P_k^{\textrm{Min}}\}_{k=1}^{K}$ are calculated sequentially in the order $k=K,K-1,...,1$. Then the proof is complete.
\end{IEEEproof}

According to Theorem 1, with the instantaneous CSI, $\{P_k^{\textrm{Min}}\}_{k=1}^{K}$ are calculated in the order $k=K,K-1,...,1$ by using (\ref{calculate Pkmin}). Afterwards, $P_{\textrm{Min}}=\sum\nolimits_{k=1}^KP_k^{\textrm{Min}}$ can be used as a threshold to verify whether $P$ is large enough to meet the constraint on data rate for each user.

\section{Energy Efficiency Maximization}\label{EEOpt}
In this section, we solve problem (\ref{EE_MV_Original}) under the condition $P\geq P_{\textrm{Min}}$, which guarantees the feasibility of problem (\ref{EE_MV_Original}).

Substituting (\ref{Rm}) into (\ref{R}), we first reformulate the achievable sum rate $R$ as follows:
\begin{equation} \label{R_expanded}
    \begin{aligned}
        &R=\log_2\left(P\left|h_{1}\right|^2\sum\nolimits_{i=1}^Ka_i+\sigma^2\right)\\
        &~~~+\sum\nolimits_{k=1}^{K-1}\left[\log_2\left(P\left|h_{k+1}\right|^2\sum\nolimits_{i=k+1}^Ka_i+\sigma^2\right)\right.\\
        &~~~~~~~~~~~~~~~~\left.-\log_2\left(P\left|h_k\right|^2\sum\nolimits_{i=k+1}^Ka_i+\sigma^2\right)\right].
    \end{aligned}
\end{equation}
For notational simplicity, we further define
\begin{subequations}\label{Simplicity}
    \begin{align}
        &C_k\triangleq P\left|h_{k}\right|^2,~~~~~~1\leq k\leq K,\\
        &\theta\triangleq\sum\nolimits_{i=1}^{K}a_i=\frac{P_t}{P},\label{definetheta}\\
        &x_k\triangleq\sum\nolimits_{i=k+1}^Ka_i,~~1\leq k\leq K-1,\label{tm}\\
        &F_k(x_k)\triangleq\log_2\left(C_{k+1}x_k+\sigma^2\right)-\log_2\left(C_{k}x_k+\sigma^2\right).\label{J_Fun}
    \end{align}
\end{subequations}
By using these notations, $R$ in (\ref{R_expanded}) is recast as
\begin{equation}\label{R_J_Fun}
    R=\log_2\left(C_1\theta+\sigma^2\right) + \sum\nolimits_{k=1}^{K-1}F_k(x_k),
\end{equation}
and the original problem (\ref{EE_MV_Original}) is rewritten as
\begin{subequations}\label{EE_MV_Theta}
    \begin{align}
    &\max_{\theta,a_k,1\leq k\leq K}~~\frac{\log_2\left(C_1\theta+\sigma^2\right) + \sum\nolimits_{k=1}^{K-1}F_k(x_k)}{\theta P+P_c}\label{EE_theta} \\
    &~~~~~\textrm{s.t.}~~~~~~~~\theta\leq1~~~\textrm{and}~~~\theta=\sum\nolimits_{k=1}^{K}a_k,\\
    &~~~~~~~~~~~~~~~~(\ref{RmQm2}).
    \end{align}
\end{subequations}
Here, we emphasize that $\theta$ is the ratio of the actually consumed transmitting power $P_t$ to the total power available at the BS $P$. In particular, $\theta$ might be less than one for maximizing the EE.
Problem (\ref{EE_MV_Theta}) can be further decoupled into two concatenate subproblems as follows.
\begin{subequations}\label{EE_MV_Theta_Decoupled}
    \begin{align}
    &\max_{\theta}~~\frac{\log_2\left(C_1\theta+\sigma^2\right) + \max\limits_{a_k,1\leq k\leq K}\sum\nolimits_{k=1}^{K-1}F_k(x_k)}{\theta P+P_c}\\
    &~~~\textrm{s.t.}~~\theta\leq1~~~\textrm{and}~~~\theta=\sum\nolimits_{k=1}^{K}a_k,\label{CosP}\\
    &~~~~~~~~(\ref{RmQm2}).\label{17c}
    \end{align}
\end{subequations}
The inner optimization problem is performed over arguments $\{a_k\}_{k=1}^K$ by taking $\theta$ as a constant, the solution of which is a function of $\theta$. Afterwards, the outer optimization problem is taken over $\theta$. These two subproblems are sequentially solved in subsections \ref{sub_Scheme} and \ref{sub_O_Theta}, respectively.
To be specific, in subsection \ref{sub_Scheme}, taking $\theta$ as a constant, we propose a power allocation strategy to solve the inner optimization problem and meanwhile obtain closed-form expressions for the optimal power allocation coefficients $\{a^*_k\left(\theta\right)\}_{k=1}^K$.
In subsection \ref{sub_O_Theta}, we prove that the outer optimization problem is a strict pseudo-concave optimization problem with respect to (w.r.t) the unique argument $\theta$, and then the bisection method is applied to find the optimal $\theta^*$ that maximizes the EE.

\subsection{Optimal Power Allocation Strategy}\label{sub_Scheme}
By fixing $\theta$ in the feasible range $\frac{P_{\textrm{Min}}}{P}\leq\theta\leq1$, the constraint $\theta\leq1$ in (\ref{CosP}) can be eliminated and then the inner optimization problem in (\ref{EE_MV_Theta_Decoupled}) is rewritten as
\begin{subequations}\label{EE_MV_Fixed_Theta}
    \begin{align}
    &\max_{a_k,1\leq k\leq K}~~\sum\nolimits_{k=1}^{K-1}F_k(x_k)\label{Ob_EE_Fixed_Theta}\\
    &~~~~\textrm{s.t.}~~~~~~~\sum\nolimits_{k=1}^{K}a_k =\theta~~~\textrm{and}~~~(\ref{RmQm2}).\label{constraintALLTheta}
    \end{align}
\end{subequations}
\begin{remark}
Actually, by regarding $\theta$ as a constant in $\frac{P_{\textrm{Min}}}{P}\leq\theta\leq1$, the nature of the inner optimization problem (\ref{EE_MV_Fixed_Theta}) is to maximize the EE subject to the constraint that the transmitting power should exactly be $\theta P$.
\end{remark}

From (\ref{EE_MV_Fixed_Theta}), we can see that the objective function in (\ref{Ob_EE_Fixed_Theta}) is the summation of $K-1$ non-convex subfunctions sharing similar forms. Based on this observation, we propose an optimization algorithm to solve (\ref{EE_MV_Fixed_Theta}), which can be elaborated in two steps as follows. \emph{Step 1}: we individually maximize each subfunction $F_k(x_k)$ subject to the constraints in (\ref{constraintALLTheta}). \emph{Step 2}: we demonstrate that the optimal solution set of each maximization problem possesses a \emph{unique common solution}. Namely, we can find a unique solution that simultaneously maximizes $F_k(x_k)$ for $1\leq k\leq K-1$ with all the constraints in (\ref{constraintALLTheta}) satisfied. Thereby, this unique solution is the optimal solution to problem (\ref{EE_MV_Fixed_Theta}).
Mathematically, denoting $\Phi_k$ as the optimal solution set for maximizing $F_k(x_k)$ subject to the constraints in (\ref{constraintALLTheta}), we will show
\begin{equation}
\Phi_{1}\cap\Phi_{2}\cap...\cap\Phi_{K-1} = \{\{a^*_i(\theta)\}_{i=1}^K\},
\end{equation}
where $\{a_i^{*}(\theta)\}_{i=1}^K$ is the unique common solution of the $K-1$ optimization problems.

\emph{Step 1}: we now solve these $K-1$ optimization problems. Firstly, the first-order derivative of $F_k(x_k)$ w.r.t $x_k$ is given as
\begin{equation}\label{lemma}
\frac{dF_k(x_k)}{dx_k}=\frac{\left(C_{k+1}-C_{k}\right)\sigma^2}{\ln2\left(C_{k+1}x_k+\sigma^2\right)\left(C_{k}x_k+\sigma^2\right)}\geq0,
\end{equation}
which demonstrates that $F_k(x_k)$ is a monotonically increasing function of $x_k$. Therefore, maximizing $F_k(x_k)$ is equivalent to maximizing $x_k$. As a result, we can uniformly formulate the aforementioned $K-1$ optimization problems as
\begin{subequations}\label{op_uniform}
    \begin{align}
    &\max_{a_k, 1\leq k\leq K}~~x_{K_0}\\
    &~~~~\textrm{s.t.}~~~~~\sum\nolimits_{k=1}^{K}a_k =\theta, \label{gamma_condition3}\\
    &~~~~~~~~~~~~~~(\ref{RmQm2}),\label{gamma_condition1}
    \end{align}
\end{subequations}
where $1\leq K_0\leq K-1$ is the index for the $K-1$ optimization problems. Problem (\ref{op_uniform}) is solved by the following proposition.

\begin{proposition}
Problem (\ref{op_uniform}) is solved when the constraints in (\ref{gamma_condition1}) are active for $1\leq k\leq K_0$, and the closed-form expressions of $\{a_k\}_{k=1}^{K_0}$ and $x_{K_0}$ are given by
\begin{subequations}\label{calculate gammatm}
    \begin{align}
    &a_k=D_k\left(\theta-\sum\nolimits_{i=1}^{k-1}a_i\right)+\frac{D_k\sigma^2}{P\left|h_k\right|^2},~~ 1\leq k\leq K_0,\label{calculate gamma}\\
    &x_{K_0}=\theta-\sum\nolimits_{k=1}^{K_0}a_k,\label{calculate tm}
    \end{align}
\end{subequations}
respectively, where $D_k=A_k/2^{R_k^{\textrm{Min}}}$.
\end{proposition}
\begin{IEEEproof}
Please see Appendix A.
\end{IEEEproof}

\emph{Step 2}: based on Proposition 1, the following theorem further gives a closed-form expression for the unique solution to problem (\ref{EE_MV_Fixed_Theta}).
\begin{theorem}\label{theo2}
The optimal power allocation coefficients $\{a^*_k\left(\theta\right)\}_{k=1}^K$ that maximize the objective function in (\ref{Ob_EE_Fixed_Theta}), are given by
    \begin{equation} \label{gammaMaxS}
        a_k^{*}(\theta)=
        \begin{cases} D_k\left(\theta-\sum\nolimits_{i=1}^{k-1}a_i^{*}(\theta)\right)+\frac{D_k\sigma^2}{P\left|h_k\right|^2},&k\neq K,\\
        \theta - \sum_{i=1}^{K-1}a_i^{*}(\theta),&k=K.
        \end{cases}
    \end{equation}
\end{theorem}
\begin{IEEEproof}
According to (\ref{calculate gamma}) in Proposition 1, arguments $\{a_k\}_{k=1}^{K_0}$ are uniquely and sequentially determined in the order $k=1,2,...,K_0$ for maximizing $x_{K_0}$. This implies that more power allocation coefficients will be determined when $K_0$ increases. Namely, the size of the optimal solution set of problem (\ref{op_uniform}), i.e., $\Phi_{K_0}$, becomes smaller as $K_0$ increases, which can be characterized by
\begin{subequations}
    \begin{align}
    &\mathfrak{}\Phi_{1}\supset\Phi_{2}\supset...\supset\Phi_{K-1},\label{setrelation1}\\
    &\Phi_{1}\cap\Phi_{2}\cap...\cap\Phi_{K-1} = \Phi_{K-1}\label{setrelation2}.
    \end{align}
\end{subequations}
Accordingly, $\Phi_{K-1}$ is the optimal solution set that simultaneously maximizes $F_{K_0}(x_{K_0})$ for $1\leq K_0\leq K-1$, consequently solving problem (\ref{EE_MV_Fixed_Theta}).
By setting $K_0$ to $K-1$ in (\ref{calculate gammatm}), the first $K-1$ optimal arguments $\{a_k^{*}(\theta)\}_{k=1}^{K-1}$ are uniquely and sequentially determined in the order $k=1,2,...,K-1$ by using (\ref{calculate gamma}). Further, we have $a_K^{*}(\theta)=\theta-\sum_{k=1}^{K-1}a_k^{*}(\theta)$ from (\ref{gamma_condition3}). As a result, the closed-form expressions of $\{a_k^{*}(\theta)\}_{k=1}^K$ that maximize the objective function in (\ref{Ob_EE_Fixed_Theta}), are given by (\ref{gammaMaxS}). Then the proof is complete.
\end{IEEEproof}

From Theorem 2, we find that the inner optimization problem (\ref{EE_MV_Fixed_Theta}) is solved when the minimum data rate constraints in (\ref{RmQm2}) are active for $1\leq k\leq K-1$, which implies that the optimal power allocation strategy is to use the extra power $\left(\theta P-P_{\textrm{Min}}\right)$ only for increasing the $K$-th user's data rate. This is because, the $K$-th user has the largest channel gain and it achieves the highest data rate among all users with the same amount of power. Namely, the $K$-th user can use power more efficiently than the other users do.
As a result, when the transmitting power is fixed as $\theta P$, the nature of maximizing the EE is to enlarge the data rate of the user with the largest channel gain as much as possible.
However, this does not signify that the extra power $\left(\theta P-P_{\textrm{Min}}\right)$ should be totally allocated to the $K$-th user, since its signal also interferes with the other $K-1$ users. More explicitly, the following corollary further reveals the essence of the proposed optimal power allocation strategy.

\begin{corollary}
$\{\frac{da^*_k(\theta)}{d\theta}\}_{k=1}^K$ are positive constants.
\end{corollary}
\begin{IEEEproof}
Please see Appendix B.
\end{IEEEproof}
Corollary 1 means that the power of the $k$-th user's signal $a^*_k(\theta)P$ increases linearly as $\theta$ increases. This implies that the extra power $\left(\theta P-P_{\textrm{Min}}\right)$ is allocated to the $k$-th user with the constant proportion $\frac{da^*_k(\theta)}{d\theta}$.

\subsection{Optimal Transmitting Power $\theta^* P$ for Maximizing the \emph{EE}}\label{sub_O_Theta}
In the previous subsection, the inner optimization problem (\ref{EE_MV_Fixed_Theta}) is solved with the closed-form solution in (\ref{gammaMaxS}), of which $\theta$ is the unique argument. Consequently, the outer optimization problem in (\ref{EE_MV_Theta_Decoupled}) is transformed into a univariate optimization problem w.r.t $\theta$, which is given by
\begin{subequations}\label{EE_SV}
\begin{align}
    &\max_{\theta}~~\frac{\log_2\left(C_1\theta+\sigma^2\right) + \sum\nolimits_{k=1}^{K-1}F_k(x_k^*(\theta))}{\theta P+P_c}\label{EE_SV_Final}\\
    &~~\textrm{s.t.}~~~P_{\textrm{Min}}\leq\theta P\leq P,\label{Feasible_Range}
\end{align}
\end{subequations}
where $x_k^*(\theta)=\sum_{i=k+1}^Ka_i^*(\theta)=\theta-\sum_{i=1}^ka_i^*(\theta)$ and the constraint in (\ref{Feasible_Range}) indicates the feasible range of $\theta$.

\begin{theorem}\label{theo3}
Denote the objective function in (\ref{EE_SV_Final}) as $\textrm{EE}(\theta)$, then EE$\left(\theta\right)$ is a strict pseudo-concave function w.r.t $\theta$.
\end{theorem}
\begin{IEEEproof}
It can be easily verified that the second-order derivative of $F_k(x_k)$ w.r.t $x_k$ is non-positive, which indicates that $F_k(x_k)$ is a concave function of $x_k$ for $1\leq k \leq K-1$. Based on this property, we further conclude that $F_k(x_k^*(\theta))$ is a concave function of $\theta$. This is because $\{a^{*}_i(\theta)\}_{i=1}^{K}$ and $\{x_k^*(\theta)\}_{k=1}^{K-1}$ are all affine mappings according to their linear expressions, which preserves the convexity of $F_k(x_k^*(\theta))$ w.r.t $\theta$. Moreover, it can be easily verified that $\log_2\left(C_1\theta+\sigma^2\right)$ is a strict concave function of $\theta$. As a result, the numerator of EE$\left(\theta\right)$, which is the summation of $\sum\nolimits_{k=1}^{K-1}F_k(x_k^*(\theta))$ and $\log_2\left(C_1\theta+\sigma^2\right)$, must be a strict concave function w.r.t $\theta$, since the convexity is preserved by the addition operation. By now, we have proved that EE$\left(\theta\right)$ has a strict concave numerator and an affine denominator, which ensures that EE$\left(\theta\right)$ is a strict pseudo-concave function w.r.t $\theta$ \cite[Proposition 6]{Pseudo-concavity}.
\end{IEEEproof}

According to Theorem 3, EE($\theta$) is a strict pseudo-concave function of $\theta$ and thus admits a unique maximizer which is the unique root of the equation $\frac{d\textrm{EE}\left(\theta\right)}{d\theta}=0$ \cite[Proposition 5]{Pseudo-concavity}. The expression of $\frac{d\textrm{EE}\left(\theta\right)}{d\theta}$ is given by (\ref{dEE}) at the top of the next page. Then, the bisection method\footnote{Problem (\ref{EE_SV}) can be also solved by Dinkelbach's algorithm or Charnes-Cooper Transform (see, e.g., \cite{EEProgram} and references therein).} can be applied to find out $\theta^*$ that maximizes EE($\theta$) with polynomial complexity.


\begin{figure*}[!t]
\normalsize
\begin{equation}\label{dEE}
    \begin{split}
    \frac{d\textrm{EE}\left(\theta\right)}{d\theta}=
    &\frac{\frac{1}{\ln2}
          \left(\frac{C_K\frac{da_K^*(\theta)}{d\theta}}{\sigma^2+C_Ka^{*}_K(\theta)}\right)\left(\theta P+P_c\right)-\left[\log_2\left(C_1\theta+\sigma^2\right) + \sum\nolimits_{k=1}^{K-1}F_k(x_k^*(\theta))\right]P}{\left(\theta P+P_c\right)^2}
    \end{split}
\end{equation}
\hrulefill
\vspace*{-6pt}
\end{figure*}

\section{Simulation Results}
In this section, we numerically evaluate the proposed energy-efficient power allocation strategy, which is labeled as ``EEPA''.
Besides, another strategy that uses full power $P$ for maximizing the SE of the system is also presented, which is labeled as ``MaxSE''. This ``MaxSE'' strategy is actually the solution of the inner optimization problem (\ref{EE_MV_Fixed_Theta}) with $\theta=1$.
For the comparison between NOMA and conventional OMA, we use a TDMA system as a baseline, where the time slots with equal duration are individually allocated to users and the transmit power is fixed, of which the maximum EE is obtained via exhausted search on the transmit power.

We solve problem (\ref{EE_MV_Original}) for 10,000 times with random channel realizations. The parameter setting is: $\{g_k\}_{k=1}^K\sim\mathcal{CN}(0,1)$, $\alpha=3$, $\sigma^2=-70$ dBm and $P_c=30$ dBm. In particular, when the total power $P$ is not large enough for guaranteeing all users' minimum required date rates, the BS will not send messages and the EE is set to zero for this case.

\begin{figure}[!t]
    \vspace{-1.0em}
    \centering
        \includegraphics[height = 5.2cm, width = 6.4cm]{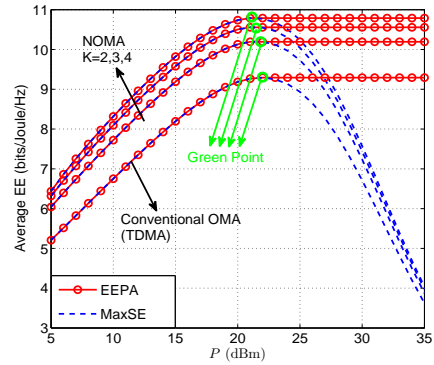}
    \vspace{-1.0em}
    \caption{Average EE (bits/Joule/Hz) versus total power available at the BS $P$ (dBm). $R_k^{\textrm{Min}}=1$ bits/s/Hz and $d_k=80$ m, where $1\leq k\leq K$.}
    \label{fig:EE_P}
\end{figure}

Fig. \ref{fig:EE_P} depicts the average EE versus $P$. We can see that there exists a ``Green Point'' at which the maximum EE is achieved by both ``EEPA'' and ``MaxSE'' strategies. When $P$ is smaller than the Green Point's corresponding power on the horizontal axis, the increase of SE will simultaneously bring an increase of EE. But when $P$ is larger, using full power $P$ is not optimal from the perspective of EE. Besides, NOMA is superior to OMA in terms of EE, and the performance gains of NOMA become more significant as $K$ increases. This is because when more users are simultaneously served, higher diversity gains and higher SE can be achieved.

\begin{figure}[!t]
    \centering
    \vspace{-1.0em}
    \includegraphics[height = 5.2cm, width = 6.4cm]{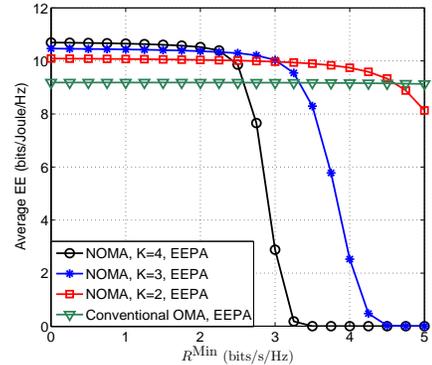}
    \vspace{-1.0em}
    \caption{Average EE versus minimum required data rate $R^{\textrm{Min}}$ for different numbers of users, where $\{R_k^{\textrm{Min}}\}_{k=1}^K = R^{\textrm{Min}}$ and $P$=20 dBm.}
    \label{fig:EE_QoS}
    \vspace{-1.0em}
\end{figure}

By setting $\{R_k^{\textrm{Min}}\}_{k=1}^K$ to the same value, denoted by $R^{\textrm{Min}}$, Fig. \ref{fig:EE_QoS} shows the average EE versus $R^{\textrm{Min}}$. We can see that as $R^{\textrm{Min}}$ increases, it is more difficult to achieve a high EE. This is because, the increase of $R^{\textrm{Min}}$ requires the BS to allocate more power to the users with worse channel conditions, which consequentially degrades the EE performance. It can be further seen that as $R^{\textrm{Min}}$ becomes very large, the EE approaches zero faster for NOMA. This is because $P$ is not large enough for satisfying the highly demanding data rate requirements and then the BS does not send messages, which implies that NOMA is more suitable for low-rate communications and less robust for the increase of data rate requirements in comparison with conventional OMA.

\begin{figure}[!t]
    \centering
    \vspace{-1.0em}
    \includegraphics[height = 5.2cm, width = 6.4cm]{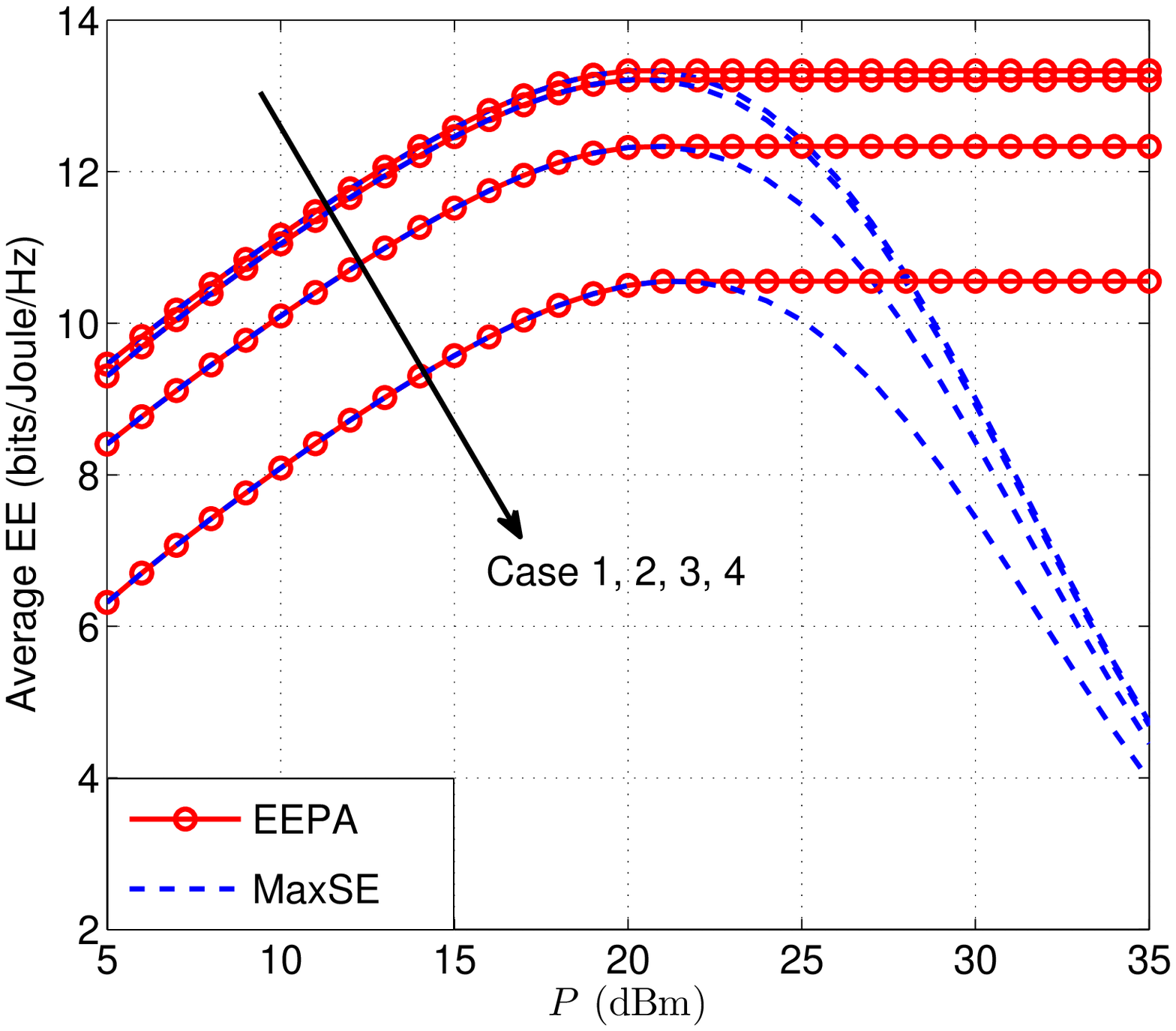}
    \vspace{-1.0em}
    \caption{Average EE (bits/Joule/Hz) versus total power available at the BS $P$ (dBm), for different cases of user locations.
    \protect\\ Case 1: $d_1$=60m, $d_2$=50m, $d_3$=40m, $(d_1+d_2+d_3)/3=50$m.
    \protect\\ Case 2: $d_1$=70m, $d_2$=55m, $d_3$=40m, $(d_1+d_2+d_3)/3=55$m.
    \protect\\ Case 3: $d_1$=60m, $d_2$=55m, $d_3$=50m, $(d_1+d_2+d_3)/3=55$m.
    \protect\\ Case 4: $d_1$=80m, $d_2$=80m, $d_3$=80m, $(d_1+d_2+d_3)/3=80$m.}
    \vspace{-1.0em}
    \label{fig:EE_Pd}
\end{figure}

Fig. \ref{fig:EE_Pd} investigates the influence of user locations on the EE. First of all, there is no doubt that the system must have a low EE when all users locate far from the BS (shown as case 4). More importantly, we can see: 1) case 1 and case 2 have very close EE; 2) case 2 outperforms case 3 although they have an equal average user distance. These observations imply that the EE performance is mainly determined by the user with the closest distance to the BS, since this user is most likely to have the largest channel gain so as to use energy most efficiently, which validates our analysis in subsection \ref{sub_Scheme}.

\section{Conclusion}
In this correspondence, we have studied the EE optimization in a SISO NOMA system where multiple users have their own data rate requirements. An energy-efficient power allocation strategy has been proposed to maximize the EE. Our numerical results have shown that NOMA has superior EE performance compared with conventional OMA. This is because, in NOMA, multiple users are simultaneously served via power domain division, which makes energy be more efficiently used.

\appendices
\section{Proof of Proposition 1}
Since problem (\ref{op_uniform}) is convex, the following KKT conditions are necessary and sufficient for the optimality of problem (\ref{op_uniform}):
\begin{align}
    &\hspace{-0.05in}\lambda =
    \begin{cases}
    \mu_k-\sum_{i=1}^{k-1}\mu_iA_i, &~~~~1\leq k \leq K_0,\\
    \mu_k-\sum_{i=1}^{k-1}\mu_iA_i+1, &~~~~K_0+1\leq k\leq K,
    \end{cases}\label{KKT_lambda}\\
    &\hspace{-0.05in}\mu_k\left[A_k\left(\sum_{i=k+1}^{K}a_i+\frac{\sigma^2}{P\left|h_{k}\right|^2}\right)-a_k\right]=0,~1\leq k\leq K,\hspace{-0.05in}\\
    &\hspace{-0.05in}\mu_k\geq 0,~~~1\leq k\leq K,\label{mu_original}
\end{align}
where $\lambda$ and $\{\mu_k\}_{k=1}^K$ are the Lagrange multipliers for constraints (\ref{gamma_condition3}) and (\ref{gamma_condition1}), respectively.
In the following, we prove that $\{\mu_k\}_{k=1}^{K_0}$ are positive numbers, which is equivalent to that the constraints in (\ref{gamma_condition1}) are active for $1\leq k\leq K_0$.

\emph{Firstly}, we demonstrate $\mu_1>0$ by contradiction: suppose $\mu_1=0$, then we have $\lambda=\mu_1=0$ by setting $k=1$ in (\ref{KKT_lambda}). Accordingly, for $1\leq k\leq K_0$ in (\ref{KKT_lambda}), we can further obtain $\mu_k=\sum_{i=1}^{k-1}\mu_iA_i$, which indicates that $\{\mu_k\}_{k=1}^{K_0}$ are all zeros, since $\mu_k=0$ can be calculated in the order $k=2,3,...,K_0$.
However, by setting $k=K_0+1$ in (\ref{KKT_lambda}), we have
\begin{equation}\label{v_t_lambda}
    \mu_1=\lambda=\mu_{K_0+1}-\sum\nolimits_{i=1}^{K_0}\mu_iA_i+1=\mu_{K_0+1}+1>0,
\end{equation}
which contradicts to the assumption that $\mu_1=0$. As a result, we have proved that $\lambda=\mu_1>0$.

\emph{Afterwards}, for $2\leq k\leq K_0$ in (\ref{KKT_lambda}), we have $\mu_k=\sum\nolimits_{i=1}^{k-1}\mu_iA_i + \lambda$ , which indicates that $\mu_k>0$ for $2\leq k\leq K_0$.
Thereby, constraints (\ref{gamma_condition1}) must be active for $1\leq k\leq K_0$.
We set constraints (\ref{gamma_condition1}) to be active for $1\leq k\leq K_0$ and replace $\sum_{i=k+1}^{K}a_i$ by $(\theta-\sum_{i=1}^{k}a_i)$ in (\ref{gamma_condition1}), then the closed-form expressions of $\{a_k\}_{k=1}^{K_0}$ and $x_{K_0}$ are derived and given by (\ref{calculate gamma}) and (\ref{calculate tm}), respectively. Specifically, $\{a_k\}_{k=1}^{K_0}$ are calculated sequentially in the order $k=1,2,...,K_0$.

\section{Proof of Corollary 1}
Firstly, $\frac{da^*_k(\theta)}{d\theta}$ can be derived from (\ref{gammaMaxS}):
\begin{equation} \label{dgammaOpt}
    \frac{da^*_k(\theta)}{d\theta} =
    \begin{cases}
    D_k\left(1-\sum_{i=1}^{k-1}\frac{da^*_i(\theta)}{d\theta}\right), &k\neq K,\\      1-\sum_{i=1}^{K-1}\frac{da^*_i(\theta)}{d\theta}, &k=K.
    \end{cases}
\end{equation}
It can be seen that $0\leq\sum_{i=1}^{k-1}\frac{da^*_i(\theta)}{d\theta}<1$ is a sufficient condition for $\frac{da^*_k(\theta)}{d\theta}>0$ due to $0\leq D_k<1$ for $1\leq k\leq K$. In the following, we use Mathematical Induction to prove
\begin{equation}\label{MI}
0\leq\sum\nolimits_{i=1}^{k-1}\frac{da^*_k(\theta)}{d\theta}<1,~~~1\leq k\leq K.
\end{equation}
It is obvious that (\ref{MI}) holds when $k=1$. When $k=N+1$,
    \begin{align}
    &\sum_{i=1}^{N}\frac{da^*_i(\theta)}{d\theta}=\sum_{i=1}^{N-1}\frac{da^*_i(\theta)}{d\theta}+D_N\left(1-\sum_{i=1}^{N-1}\frac{da^*_i(\theta)}{d\theta}\right)\nonumber\\
    &~~~~~~~~~~~~~~=(1-D_N)\sum\nolimits_{i=1}^{N-1}\frac{da^*_i(\theta)}{d\theta}+D_N.\label{MIK}
    \end{align}
By using the induction hypothesis that (\ref{MI}) holds when $k=N$, i.e., $0\leq\sum_{i=1}^{N-1}\frac{da^*_i(\theta)}{d\theta}<1$, we have $0\leq\sum_{i=1}^{N}\frac{da^*_i(\theta)}{d\theta}<1$. Thereby, we have proved $\frac{da^*_k(\theta)}{d\theta}>0$. Besides, $\{\frac{da^*_k(\theta)}{d\theta}\}_{k=1}^K$ are calculated sequentially in the order $k=1,2,...,K$.

\end{document}